# Traffic Incident Analysis on Urban Arterials Using Extended Spectral Envelope Method

Zhen-zhen Yang [a,c], Liang Gao [a,*], Zi-you Gao [a], Ya-fu Sun [b,c], Sheng-min Guo [c,d]

[a] State Key Laboratory of Rail Traffic Control and Safety, MOE Key Laboratory for Urban Transportation Complex Systems Theory and Technology, Systems Science Institute, Center of Cooperative Innovation for Beijing Metropolitan Transportation, Beijing Jiaotong University, Beijing 100044, China

[b] Research and Development Center on Intelligent Transport Technology & Equipment, Ministry of Transport, Beijing 100191, China

[c] Beijing PalmGo Information and Technology Co., Ltd, Beijing 100191, China

[d] State Key Laboratory of Software Development Environment, Beihang University, Beijing 100191, China

**ABSTRACT**

A traffic incident analysis method based on extended spectral envelope (ESE) method is presented to detect the key incident time. Sensitivity analysis of parameters (the length of time window, the length of sliding window and the study period) are discussed on four real traffic incidents in Beijing. The results show that: (1) Moderate length of time window got the best accurate in detection. (2) The shorter the sliding window is, the more accurate the key incident time are detected. (3) If the study period is too short, the end time of an incident cannot be detected. Empirical studies show that the proposed method can effectively discover the key incident time, which can provide a theoretic basis for traffic incident management.

**Key words**: traffic incident; extended spectral envelope; key incident time; traffic phase transition

## 1. Introduction

Effective traffic incident management (TIM) can retrieve the loss caused by traffic incident (Baiocchi et al., 2015; Farradyne, 2000; Hou et al., 2013; Li et al., 2015; Long et al., 2012), such as decreasing travel time, air pollution, and fuel consumption. Determining the key incident time is important for a TIM system.

Researchers have applied many statistical methods to analyze the key incident time, such as Bayesian decision tree (Jiyang et al., 2008), probabilistic distribution analysis (Gobol et al.,1987), linear regression analysis (Garib et al.,1997; Wang, et

[*] Corresponding author: Tel.: +86 (0)10 51687143.
   E-mail addresses: yangzhenzhen@bjtu.edu.cn (Z.Z. Yang), lianggao@bjtu.edu.cn (L. Gao), zygao@bjtu.edu.cn (Z.Y. Gao) , sunyafu@chinatransinfo.com(Y. F. Sun), guosm@nlsde.buaa.edu.cn (S.M. Guo).

al.,2008), conditional probability analysis (Nam and Mannering, 2000), time-sequential model (Khattak et al.,1995), M5P tree algorithm (Zhan et al., 2011), fully parametric hazard-based duration models (Abdulla et al.,2011). These methods neglect the actual traffic flow characteristics and their accuracy has much room for improvement.

Traffic incident usually generates bottleneck, which may arise traffic oscillations (Nie, 2010) and induce traffic phase transition (Daganzo et al., 1999). So, the key incident time can be considered as the time when traffic oscillation or traffic phase transition happens. Analyzing traffic oscillations and traffic phase transition is a promising way to improve the detection of key incident time. Existing studies can be categorized as oblique cumulative curves (Muñoz and Daganzo, 2003; Sarvi et al., 2007), wavelet transform (Jeong et al., 2011; Zheng, 2011), describing-function approach (DFA) (Li et al., 2011; 2012; 2014), frequency spectrum analysis approach (Li et al., 2010), extended spectral envelope (ESE) method (Zhao et al., 2014).

Typically, traffic phase transition was detected by raw data, aggregated data, or oblique cumulative curves (Muñoz and Daganzo, 2003; Sarvi et al., 2007). In these analyses, transition was identified at the time when congestion occurred. Because of noise and data resolution, identifying precise incident time is subjective.

Wavelet transform was employed by Zheng et al. (2011) to analyze the features related to bottleneck activities and traffic oscillations in a systematic manner. Furthermore, the origins and the microscopic features of stop-and-go wave were identified by analyzing vehicle trajectories.

Describing-function approach (DFA) was adopted by Li et al. (2011) to analytically predict the properties of oscillation propagation for a general class of nonlinear car-following laws. Later, DFA method in quantification of traffic oscillations was validated by real traffic data (Li et al. 2012), and its potential to explain oscillation-related empirical traffic studies, such as estimating the fuel consumption and emission during traffic oscillations, was explored extensively (Li et al. 2014).

In addition, Li et al. (2010) proposed a frequency spectrum analysis approach to improve the measurements of traffic oscillation properties (e.g. periodicity, magnitude) from field data. The useful oscillation information was distinguished by a standard signal processing technique from noise and non-stationary traffic trends. Driver-experienced oscillations were estimated via detector data.

ESE method was proposed by Zhao et al. (2014) to analyze traffic oscillations and to reveal detailed spatial-temporal profiles of traffic oscillations. ESE method was the combination of long-term and short-term spectral envelope analysis. Long-term spectral envelope analysis computed spectral envelope of all time windows at all locations. Short-term spectral envelope analysis divided the study period into relatively short time windows and carried the spectral envelope analysis in each of windows separately. The advantage of short-term spectral envelope analysis is to capture the time-varying properties.

Some other researches have been done to analyze the characteristics of oscillations and phase transitions. Such as Chen et al. (2014) proved that periodicity and development of traffic oscillations. Blandin et al. (2013) proposed a phase transition model of non-stationary traffic in conservation form. The space gap between two vehicles would oscillate around the desired space gap in the deterministic limit in congested traffic flow (Tian et al., 2015).

This paper utilizes ESE method to capture the time-varying properties and to determine the key incident time. The rest of this paper is organized as follows. Section 2 introduces the ESE method and the method on how to verdict the key incident time. In Section 3 and 4, sensitivity analysis of parameters (the length of time window, the length of sliding window and the study period) are discussed on four real traffic incidents in Beijing. Section 5 concludes the paper.

## 2. ESE method

Extended Spectral Envelope (ESE) is an extended method based on spectral envelope (SE). Spectral envelope (SE) method aims to emphasize salient oscillations by selecting certain transformations for the entire study period, which cannot be used to detect the start and end time of an oscillation at a specific frequency, and not to investigate spatial-temporal evolution of traffic and transportation system (Zhao et al., 2014). ESE is proposed to capture such dynamic properties. The idea of ESE is to divide the study period into short windows with some overlapping and compute SE for each window in a sequential order.

$T$ is the study period, and is divided into short time windows, e.g. 5 minutes window. Define $L$ as the length of time window, $\eta$ as the overlap between the adjacent time windows; $\delta$ as the length of sliding window. The relationship among them is shown in Figure 1.

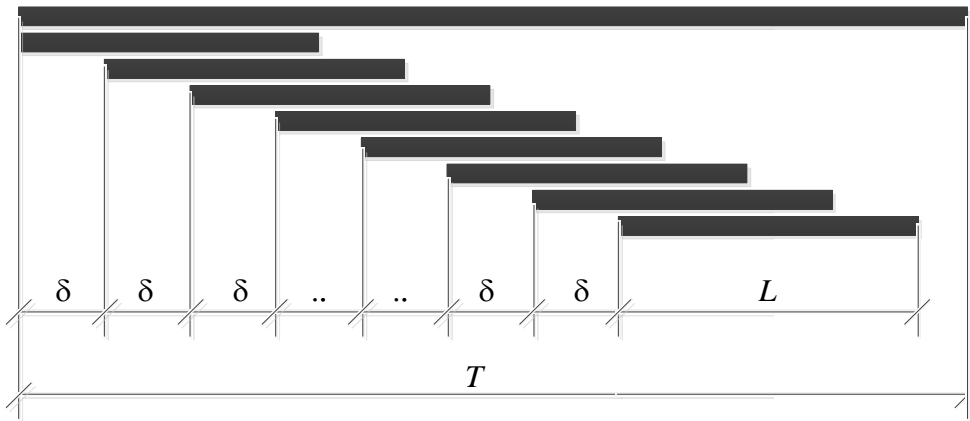

Figure 1. The relationship among the study period ($T$), the length of time window ($L$), and the length of sliding window ($\delta$).

In ESE method, the time and space arguments are discredited. Define $m$ as the number of all analyzed links, $n$ as the number of time series, $V$ as the velocity set of time series data.

$$V = \begin{bmatrix} v_1(1) & \cdots & v_1(n) \\ \vdots & \ddots & \vdots \\ v_m(1) & \cdots & v_m(n) \end{bmatrix}$$

$$v_i = [v_i(1), v_i(2), \ldots, v_i(n)]$$

Let $\varepsilon$ represent the ID of a time window, $N$ is the total number of time windows. i.e. $\varepsilon = 1, 2, \ldots, N$. The truncated time series at all location then form a truncated matrix.

$$V^\varepsilon = \begin{bmatrix} v_1^\varepsilon(1) & \cdots & v_1^\varepsilon(n) \\ \vdots & \ddots & \vdots \\ v_m^\varepsilon(1) & \cdots & v_m^\varepsilon(n) \end{bmatrix}$$

$$v_i^\varepsilon = [v_i^\varepsilon(1), v_i^\varepsilon(2), \ldots, v_i^\varepsilon(n)]$$

All time windows form the following matrix

$$\overline{V} = \begin{bmatrix} V^1 \\ V^2 \\ \vdots \\ V^N \end{bmatrix}$$

Then spectral envelope $\lambda^\varepsilon(\omega)$ and the optimal weight vector $\beta^{\omega,\varepsilon}$ of time window $\varepsilon$ are calculated by ESE method.

The spatial covariance $c_{i,j}^\varepsilon$ and correlation $r_{i,j}^\varepsilon$ between time series $v_i^\varepsilon$ and $v_j^\varepsilon$ are given by

$$c_{i,j}^\varepsilon = E[(v_i^\varepsilon - \mu_i^\varepsilon)(v_j^\varepsilon - \mu_j^\varepsilon)^\Phi]$$

$$r_{i,j}^\varepsilon = \frac{c_{i,j}^\varepsilon}{\sigma_i^\varepsilon \sigma_j^\varepsilon}$$

where the symbol "$\Phi$" denotes the conjugate transpose of the vector; $E[\cdot]$ is the expectation operator; $\mu_i^\varepsilon$ and $\sigma_i^\varepsilon$ are the mean and standard deviation of the time series $v_i^\varepsilon$, respectively. $C_V^\varepsilon$ and $R_V^\varepsilon$ denoted the covariance and correlation matrices associated with $V^\varepsilon$.

The cross-power spectral density $p_{i,j}^\varepsilon(\omega)$ measures the concentration of cross-power of the time-series at a given frequency ω. $p_{i,j}^\varepsilon(\omega)$ is obtained through a Fourier transform of the cross-covariance function of time series $v_i^\varepsilon$ and $v_j^\varepsilon$.

$$p_{i,j}^\varepsilon(\omega) = \sum_{\tau=-T}^{T} g_{i,j}^\varepsilon(\tau) \cdot e^{-j\omega\tau}$$

where ω is the Fourier frequency defined as cycle per unit time interval and ranges between $-1/2$ and $1/2$. The cross-covariance function $g_{i,j}^\varepsilon(\tau)$ measures the covariance between the two time series with a time lag $\tau$.

$$g_{i,j}^\varepsilon(\tau) = E[(v_i^\varepsilon - \mu_i^\varepsilon)(v_j^{\varepsilon,\tau} - \mu_j^\varepsilon)^\Phi]$$

$$v_j^{\varepsilon,\tau} = [v_i^\varepsilon(1+\tau), v_i^\varepsilon(2+\tau), \dots, v_i^\varepsilon(n+\tau)]$$

The cross power spectral density matrix for $V^\varepsilon$ at the frequency ω is denoted as

$$P_V^\varepsilon(\omega) = \begin{bmatrix} p_{1,1}^\varepsilon(\omega) & \cdots & p_{1,m}^\varepsilon(\omega) \\ \vdots & \ddots & \vdots \\ p_{m,1}^\varepsilon(\omega) & \cdots & p_{m,m}^\varepsilon(\omega) \end{bmatrix}$$

The spectral envelop of $V^\varepsilon$ at $\omega \in [-1/2, 1/2]$ is defined as

$$\lambda^\varepsilon(\omega) := \sup_{\beta^\varepsilon \neq 0} \left\{ \frac{P_S^\varepsilon(\omega, \beta^\varepsilon)}{C_S^\varepsilon(\beta^\varepsilon)} \right\} = \sup_{\beta^\varepsilon \neq 0} \left\{ \frac{(\beta^\varepsilon)^\Phi P_V^\varepsilon(\omega) \beta^\varepsilon}{(\beta^\varepsilon)^\Phi C_V^\varepsilon \beta^\varepsilon} \right\}$$

where $\beta^\varepsilon$ is a complex vector that can not only scale the time series but also offset the time lags. $C_S^\varepsilon(\beta^\varepsilon)$ is the total power of the weighted time series. $C_S^\varepsilon(\beta^\varepsilon) = (\beta^\varepsilon)^\Phi C_V^\varepsilon \beta^\varepsilon = \int_{-1/2}^{1/2} P_S^\varepsilon(\omega, \beta^\varepsilon) d\omega = 2\int_0^{1/2} P_S^\varepsilon(\omega, \beta^\varepsilon) d\omega$. $P_S^\varepsilon(\omega, \beta^\varepsilon)$ is the power concentrated at the frequency ω, corresponding to the linear transformation defined by $\beta^\varepsilon$. $P_S^\varepsilon(\omega, \beta^\varepsilon) = (\beta^\varepsilon)^\Phi P_V^\varepsilon(\omega) \beta^\varepsilon = P_S^\varepsilon(-\omega, \beta^\varepsilon)$.

The amplitude of $\lambda^\varepsilon(\omega)$ represents the largest portion of power that can be obtained from all transformed frequency ω. To evaluate the spectral envelope, let $\varphi^\varepsilon \equiv (C_V^\varepsilon)^{1/2} \beta^\varepsilon$ be an eigenvector of a matrix

$$D_V^\varepsilon \equiv (C_V^\varepsilon)^{-1/2} P_V^\varepsilon (C_V^\varepsilon)^{1/2}$$

where its corresponding eigenvalue is $\frac{(\beta^\varepsilon)^\Phi P_V^\varepsilon(\omega)\beta^\varepsilon}{(\beta^\varepsilon)^\Phi C_V^\varepsilon \beta^\varepsilon}$.

The spectral envelope equals the largest eigenvalue of matrix $D_V^\varepsilon$, corresponding to an eigenvector

$$\varphi^\varepsilon = (C_V^\varepsilon)^{1/2}\beta^\varepsilon.$$

When most of the elements of $\varphi^\varepsilon$ change from positive to negative, a sudden phase transition of traffic flow happens at time window ε. Therefore, we find out the key incident time by detecting the sudden change of $\varphi^\varepsilon$ for each consecutive window.

## 3. Data description

The data used in this empirical study were collected from operating taxis in Beijing. Four incidents in Beijing are analyzed by ESE method in our empirical study. Their locations and study period are showed in Table 1. The solid arrows in Figure 2 show the direction of traffic.

For each incident, data are aggregated every 5 minutes as a time unit. Incident #1, #2 and #4 happened on the 4th Ring Road. Incident #3 happened on the 2th Ring Road.

**Table 1**
The location and study period of four incidents.

| Incident | Location | Study period | Date | Time Unit(s) |
|---|---|---|---|---|
| #1 | Siyuan Bridge | 05:00 to 24:00 | 2012-09-07 | 229 |
| #2 | Majialou Bridge | 17:00 to 18:50 | 2012-03-01 | 23 |
| #3 | Kaiyang Bridge | 16:35 to 18:30 | 2012-03-01 | 24 |
| #4 | Wukesong Bridge | 13:15 to 18:15 | 2012-03-01 | 61 |

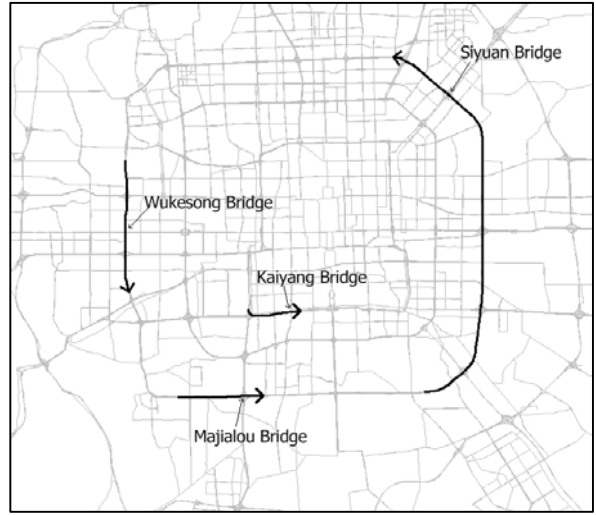

Figure 2. The empirical study area.

## 4. Empirical results

Figure 3(a ~ d) represent the speed of four incidents, respectively. Incident #1 is the most serious incident. To study the impact of $T$, we specifically choose a much longer $T$ for incident #1. $T$ is shorter for incident #2, #3, #4. The link numbers of incident #1, #2, #3, #4, are respectively 56, 10, 7, and 18. $T$ contains the incident duration and some non-incident period.

In Figure 3(a), the duration time of incident #1 is from 20 to 60 (in dark blue). Due to the evening rush hours, traffic becomes congestion and the second valley appears again. The free traffic at begin and end in Figure 3(a) respectively denote the early morning and the night.

In Figure 3(d), the duration time of incident #4 is from 30 to 50 (in dark blue). In the end of Figure 3(d), the speed is still low (in light blue), but not recovers to free. The reason for this phenomenon is same to incident #1: the evening rush hours arrive and traffic sank into normal slow.

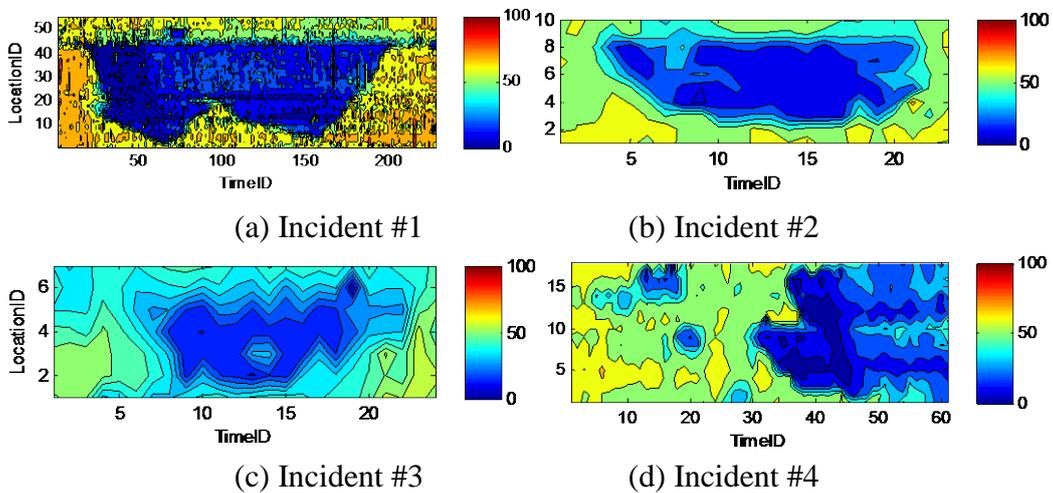

(a) Incident #1  (b) Incident #2

(c) Incident #3  (d) Incident #4

Figure 3. The speed of four incidents.

The ESE method is used to detect the key incident time. To demonstrate the impact of $L$ and $\delta$, we analyze different settings of them for each incident, as shown in Table 2.

**Table 2**
The different settings of $L$ and $\delta$ for each incident.

| Incident | $L$ | $\delta$ |
|---|---|---|
| #1 | 10,20,50,74,115,150,200,201 | 1 |
|  | 115 | 1,5,10,20 |
|  | 200 | 1,4,5,10 |
| #2 | 5,10,12,15,19,20 | 1 |
|  | 19 | 1,2,3,4 |
| #3 | 5,10,12,15,19,20 | 1 |
|  | 15 | 1,2,3,4 |
| #4 | 5,10,20,30,40,50 | 1 |
|  | 20 | 1,2,3,4 |

### 4.1. Incident #1

For incident #1, $\delta = 1$ time unit, eight scenarios, with $L = \{10,20,50,74,115,150,200,201\}$, respectively, are numerically investigated (Figure 4 and 5); $L = 115$ time units, four scenarios, with $\delta = \{1, 5, 10, 20\}$ respectively, are numerically investigated (Figure 6); $L = 200$ time units, four scenarios, with $\delta = \{1, 4, 5, 10\}$ respectively, are numerically investigated (Figure 7).

In Figure 4(a), there are negative values for the elements of $\varphi^\varepsilon$ in the 20th time window (at 6:35 am) and the 60th time window (at 9:55 am), which means two phase transitions happened in the traffic flow. For example, the 30th element of $\varphi_{30}^{19}$ is positive; the 30th element of $\varphi_{30}^{20}$ is negative; this means a sudden change of traffic flow in the 20th time window at link 30. Before 6:35 am, the traffic flow gradually increases. At 6:35 am, the negative elements are observed (Figure 4(b)), which means the traffic flow is suddenly influenced by some event such as a traffic incident. In the 60th window at 9:55 am, the negative elements are observed again (Figure 4(c)). Taking the observed phase transition in the 20th window, we can judge that the queue length of the traffic jam caused by a traffic incident begins shrinking.

In Figure 5, $\delta = 1$ time unit, eight scenarios, with $L = \{10,20,50,74,115,150,200,201\}$, respectively, are numerically investigated. While $L = \{10,20,50\}$, the significant vertical lines are too dense, causing that the key incident time cannot be obtained. With the increase of $L$, the significant vertical lines appear clearer. But when $L = 201$, there is no significant vertical line. The phenomenon states that $L$ cannot be too short or too long.

As shown in Figure 6, the only $\delta = 1$ can dig out the two significant vertical lines, where $= 115$ , which corresponds to the start and end time points respectively.

In Figure 7, both $\delta = \{1,4\}$ can dig out one significant vertical lines, but $\delta = 1$ is much clearer. All scenarios prove that $\delta$ dramatically influences the accuracy of the detection of key incident time. The smaller $\delta$ is, the more accurate the key incident time is detected.

Hence, the clearly significant vertical lines can only be achieved with appropriate value of $L$ and $\delta$.

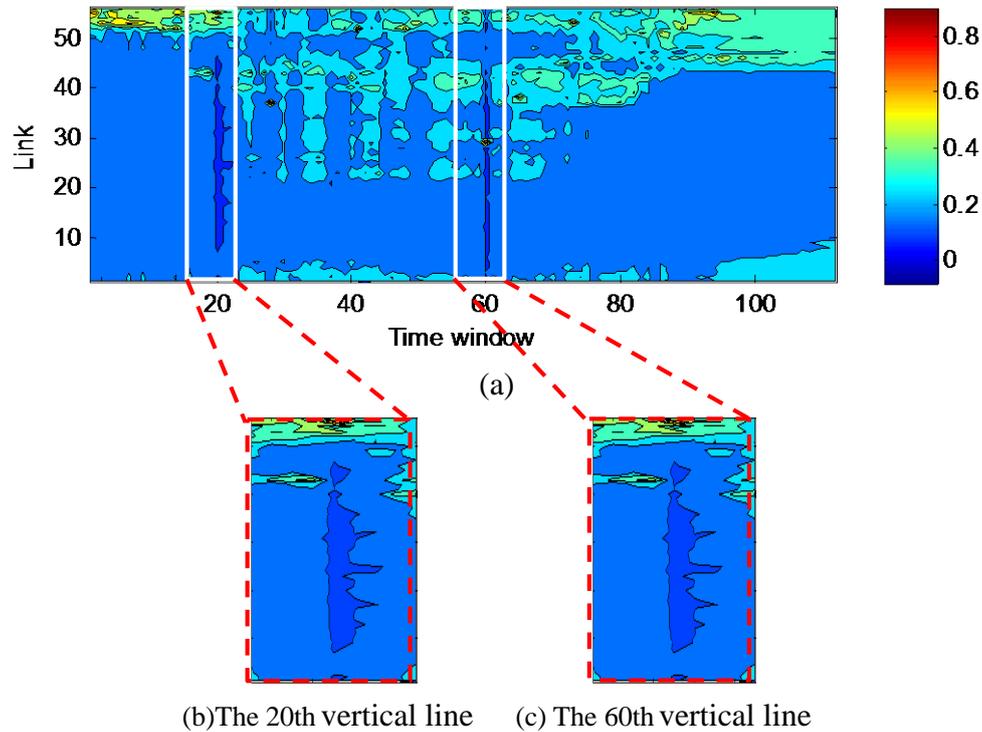

(b) The 20th vertical line   (c) The 60th vertical line

Figure 4. Incident #1: The contour plot of the elements of the eigenvector $\varphi^\varepsilon$ for each time window $\varepsilon$, where $L = 115, \delta = 1$. The elements of the eigenvector $\varphi^{20}$ in (b) and $\varphi^{60}$ in (c) are negative (colored in dark blue). Two phase transitions of traffic flow happen at the 20th and the 60th windows, respectively.

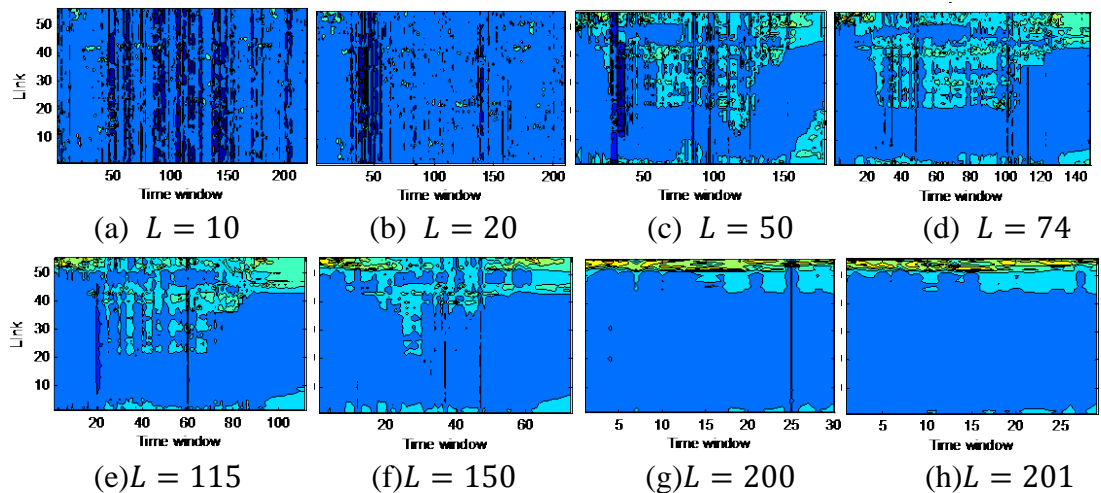

(a) $L = 10$    (b) $L = 20$    (c) $L = 50$    (d) $L = 74$

(e) $L = 115$    (f) $L = 150$    (g) $L = 200$    (h) $L = 201$

Figure 5. Incident #1: The contour plot of the elements of the eigenvector $\varphi^\varepsilon$ for each time window $\varepsilon$, where $\delta = 1$.

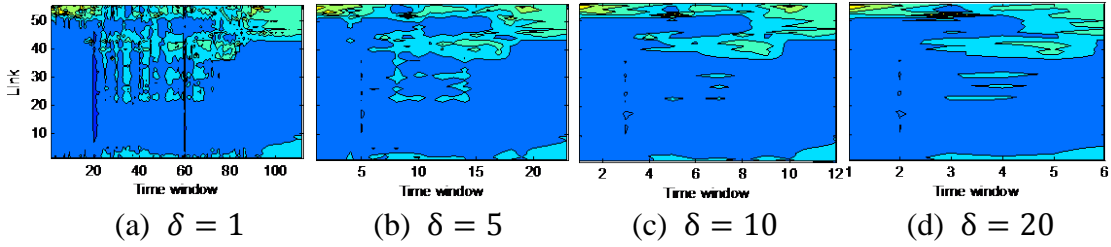

(a) $\delta = 1$  (b) $\delta = 5$  (c) $\delta = 10$  (d) $\delta = 20$

Figure 6. Incident #1: The contour plot of the elements of the eigenvector $\varphi^\varepsilon$ for each time window $\varepsilon$, where $L = 115$.

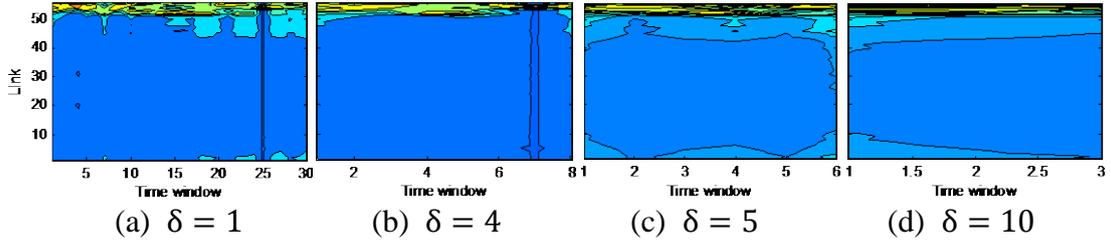

(a) $\delta = 1$  (b) $\delta = 4$  (c) $\delta = 5$  (d) $\delta = 10$

Figure 7. Incident #1: The contour plot of the elements of the eigenvector $\varphi^\varepsilon$ for each time window $\varepsilon$, where $L = 200$.

### 4.2. Incident #2

For incident #2, $\delta = 1$ time unit, six scenarios, with $L = \{5,10,12,15,19,20\}$, respectively, are numerically investigated (Figure 8); $L = 19$ time units, four scenarios, with $\delta = \{1, 2, 3, 4\}$ respectively, are numerically investigated (Figure 9).

In Figure 8(a), too many negative values for the elements of $\varphi^\varepsilon$ are detected, which are not the key incident time and we cannot recognize key information in the chaos state. With the increasing value of $L$, the chaos is eliminated gradually. Until $L = 19$, the incident start time can be unambiguously detected without any interference(Figure 8(e)). But when $L = 19$ pluses 1 to $L = 20$, the significant vertical line disappears with no transition (Figure 8(f)). Taking the observed phase transition in the 3th window, we can judge that the queue length of the traffic jam caused by a traffic incident begins. The phenomenon also states that $L$ cannot be too short or too long.

As shown in Figure 9, both $\delta = \{1,2\}$ can dig out a significant vertical line, where $= 19$, corresponding to the start time point. With the in-depth comparison, $\delta = 1$ is much more concentrated. There is no vertical line while $\delta = \{3,4\}$, which verify the results of incident #1 : the smaller $\delta$, the more accurate the key incident time is detected.

The phenomenon in incident #2 emphasizes that, the clearly significant vertical lines can only be achieved with appropriate value of $L$ and $\delta$.

In incident #2, the reason, we cannot obtain the end time of incident, is that $T$ is not long enough and the appropriate time window cannot slide to the end time point. As shown in Figure 3(b), the incident end time is approximately at 21th time unit; the whole analysis time units are 23. If we force to exhibit the incident end time unit, the time window must be set no more than 2 time units. But the study result above proves that, there are too much chaos with too short length of time window. So in incident #2, we cannot obtain the incident end time with these 23 time units data.

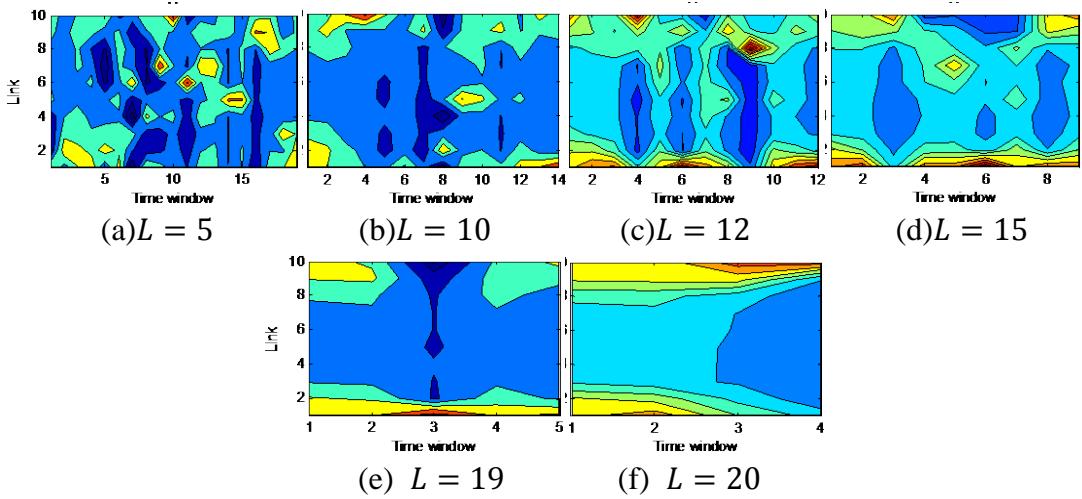

(a) $L = 5$  (b) $L = 10$  (c) $L = 12$  (d) $L = 15$

(e) $L = 19$  (f) $L = 20$

Figure 8. Incident #2: The contour plot of the elements of the eigenvector $\varphi^\varepsilon$ for each time window $\varepsilon$, where $\delta = 1$.

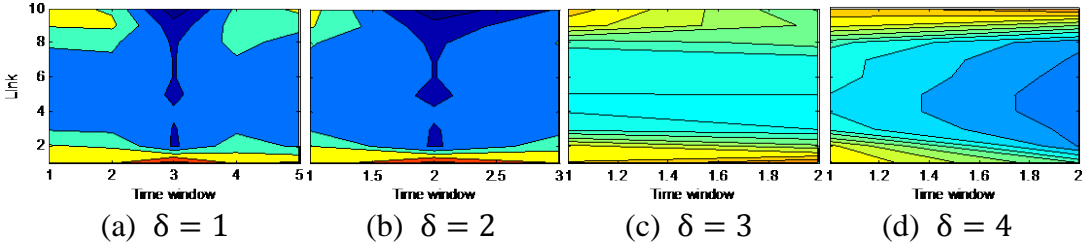

(a) $\delta = 1$  (b) $\delta = 2$  (c) $\delta = 3$  (d) $\delta = 4$

Figure 9. Incident #2: The contour plot of the elements of the eigenvector $\varphi^\varepsilon$ for each time window $\varepsilon$, where $L = 19$.

### 4.3. Incident #3

For incident #3, $\delta = 1$, six scenarios, with $L = \{5,10,12,15,19,20\}$, respectively, are numerically investigated (Figure 10); $L = 15$ time units, four scenarios, with $\delta = \{1, 2, 3, 4\}$ respectively, are numerically investigated (Figure 11).

In Figure 10(a), $L = 5$, too many negative values for the elements of $\varphi^\varepsilon$ are detected, which are not the key incident time and we cannot recognize key information in this chaos state. With the increasing value of $L$, the chaos is eliminated gradually. Only two significant vertical lines remain with less interference when $L = 12$. From Figure 3(c), we can see that the incident start time is detected by the 8th time window. In Figure 10(e ~ f), $L = \{19,20\}$, there are no negative values for the elements of the eigenvector $\varphi^\varepsilon$, verifying the result that $L$ cannot be too short or too long.

As shown in Figure 11, only $\delta = 1$ can dig out the several significant vertical line, where $L = 15$. There is no vertical line while $\delta = \{2,3,4\}$, verifying the results of incident #1 and #2 : the smaller δ, the more accurate the key incident time is detected.

Similar to the incident #2, the end time of incident also cannot be obtained in incident #3, due to the short $T$.

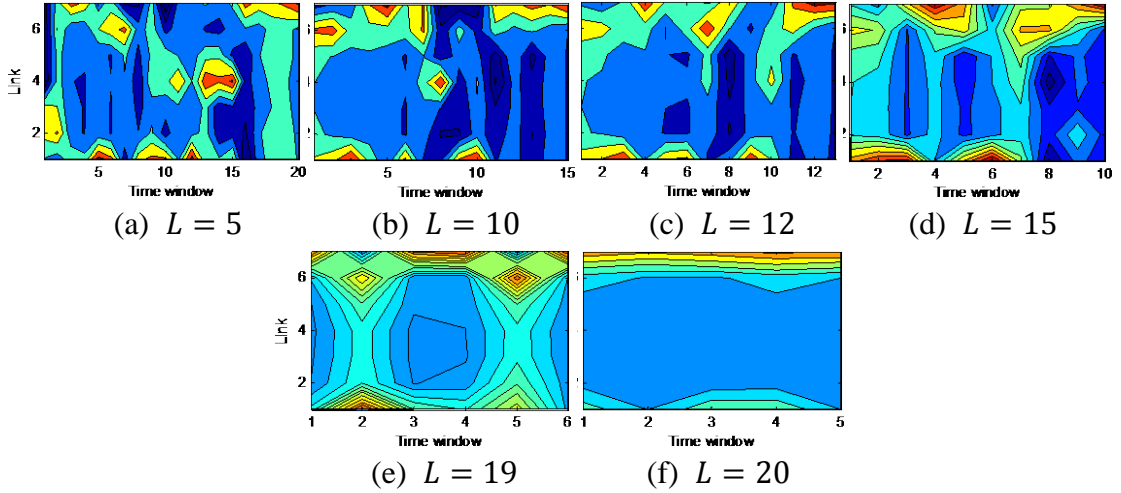

Figure 10. Incident #3: The contour plot of the elements of the eigenvector $\varphi^\varepsilon$ for each time window $\varepsilon$, where $\delta = 1$.

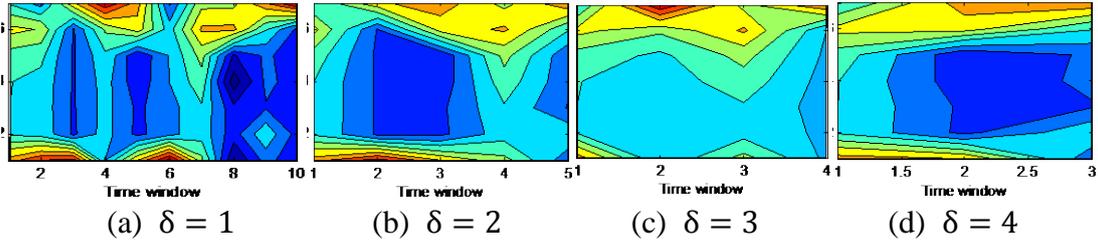

Figure 11. Incident #3: The contour plot of the elements of the eigenvector $\varphi^\varepsilon$ for each time window $\varepsilon$, where $L = 15$.

### 4.4. Incident #4

For incident #4, $\delta = 1$ time unit, six scenarios, with $L = \{5,10,20,30,40,50\}$, respectively, are numerically investigated (Figure 12). $L = 20$ time units, four scenarios, with $\delta = \{1,2,3,4\}$ respectively, are numerically investigated (Figure 13).

In Figure 12(a), $L = 5$, too many negative values for the elements of $\varphi^\varepsilon$ are detected, which are not the key incident time and we cannot recognize key information in this chaos state. With the increasing value of $L$, the chaos is eliminated gradually. When $L = \{10,20,30\}$, the significant vertical lines appear. But with the increasing of $L = \{40,50\}$, the significant vertical lines become more and more indistinct and the negative values disappear (Figure 12(e ~ f)).

In Figure 13, $\delta = \{1,2\}$ can dig out the significant vertical lines (Figure 13(a ~ b)), where $L = 20$. The vertical lines become indistinct while $\delta = \{3,4\}$ (Figure 13(c ~ d)). These results verify the results of incident #1, #2 and #3 : the smaller δ, the more accurate the key incident time is detected.

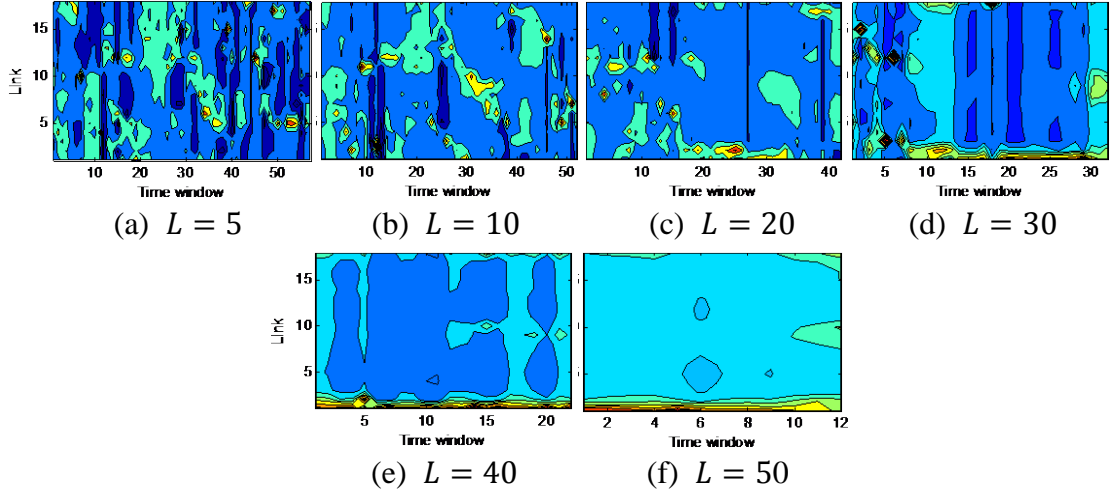

Figure 12. Incident #4: The contour plot of the elements of the eigenvector $\varphi^{\varepsilon}$ for each time window $\varepsilon$, where $\delta = 1$.

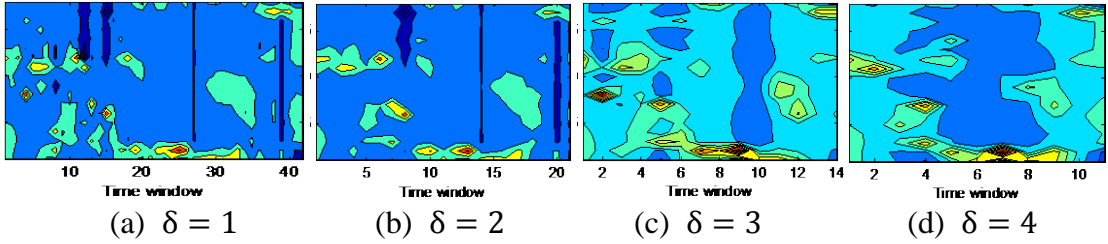

Figure 13. Incident #4: The contour plot of the elements of the eigenvector $\varphi^{\varepsilon}$ for each time window $\varepsilon$, where $L = 20$.

## 5. Discussion

Four case studies show that the length of time window $L$ is a critical parameter for ESE method in detecting the key incident time of traffic incident. If $L$ is set too short, too many negative elements of $\varphi^{\varepsilon}$ are created, and the key incident time cannot be accurately identified. With the increasing value of $L$, the chaos state is eliminated gradually and key incident time will appear. If $L$ is too long, the significant vertical lines will disappear.

To study the impact of the length of sliding window $\delta$, we fix $L$. Four empirical studies prove: the smaller $\delta$, the more accurate the key incident time is detected. In the majority situation, $\delta = 1$ can obtain the best results. With the increasing of $\delta$, the significant vertical lines gradually become more and more blurred and fade away in the end.

Furthermore, the study period $T$ influences the detection results of key incident time. In incident #2 and #3, the reason, we cannot obtain the end time of incident, is that $T$ is not long enough and the time window cannot slide to the end time point.

In general, the clearly significant vertical lines can only be achieved with appropriate value of $L$ and $\delta$. The sensitivity analysis of parameters shows that: (1) Moderate length of time window got the best accurate in detection. (2) The shorter the sliding window is, the more accurate the key incident time are detected. (3) If the study period is too short, the end time of an incident cannot be detected.

## 6. Conclusion

This paper employs the extended spectral envelope (ESE) method to deal with the traffic incident analysis. Real traffic data analyses prove that the ESE method can be used to detect the key incident time by decrypting the phase transitions of the traffic.

Sensitivity analysis of parameters (the length of time window, the length of sliding window and the study period) are discussed on four real traffic incidents in Beijing. The results show that: (1) Moderate length of time window got the best accurate in detection. (2) The shorter the sliding window is, the more accurate the key incident time are detected. (3) If the study period is too short, the end time of an incident cannot be detected.

All results indicating that the ESE method is providing a promising tool for the traffic incident analysis, and will foster the applications of traffic incident management system in ITS.

**Acknowledgement**
We thank the financial support from the Major State Basic Research Development Program of China (973 Program) No. 2012CB725400, the National High Technology Research and Development Program of China (863 Program) No. 2015AA124103, the National Natural Science Foundation of China (71101009, 71131001), the Fundamental Research Funds for the Central Universities No.2015JBM058. This work is partially supported by the State Key Laboratory of Rail Traffic Control and Safety (Contract No. RCS2014ZTY8), Beijing Jiaotong University.